\documentstyle[12pt]{article}

\def\[{\begin{eqnarray}}
\def\]{\end{eqnarray}}
\def\non{\nonumber}

\begin{document}
\pagestyle{empty}

\begin{flushright}
{\bf UA/NPPS-1-99}\\
{\bf McGILL-99-14}
\end{flushright}

\vglue 2cm
\begin{center}\begin{Large}\begin{bf}
Polarized photoproduction of large-$p_T$ hadron pairs as a probe
of the polarized gluon distribution
\end{bf}\end{Large}\end{center}
\vglue 0.35cm
{\begin{center} 
G. Grispos $^{a,1}$, A. P. Contogouris $^{a,b,2}$ and G. Veropoulos$^{a,3}$ 
\end{center}}
\parbox{6.4in}{\leftskip=1.0pc
{\it a.\ Nuclear and Particle Physics, University of Athens, 
Athens 15771, Greece }\\
\vglue -0.25cm
{\it b.\ Department of Physics, McGill University, Montreal
H3A2T8, Canada}\\}
\begin{center}
\vglue 0.5cm
Revised version
\vglue 0.5cm
\begin{bf} ABSTRACT \end{bf}
\end{center}

Production of large-$p_T$ hadron pairs by polarized photon on a longitudinally
polarized proton towards probing the polarized gluon distribution is studied.
Resolved photon contributions and the effect of changing the scales are taken
into account, and predictions are presented. A very recent experimental result
at c.m. energy $7.18$ $GeV$ is compared to our predictions extented down to this
energy. A proper combination of cross sections is also considered.

\renewcommand{\thefootnote}{\arabic{footnote}}
\addtocounter{footnote}{1}
\footnotetext{e-mail: ggrispos@cc.uoa.gr}
\addtocounter{footnote}{1}
\footnotetext{e-mail: acontog@cc.uoa.gr, apcont@physics.mcgill.ca}
\addtocounter{footnote}{1}
\footnotetext{e-mail: gverop@cc.uoa.gr}

\newpage
\pagestyle{plain}
\setcounter{page}{1}

\renewcommand{\theequation}{1.\arabic{equation}}
\begin{center}\begin{large}\begin{bf}
I. INTRODUCTION
\end{bf}\end{large}\end{center}
\vglue .2cm 

The determination of the size and shape of the polarized gluon distribution $
\Delta g$ remains a major problem in Spin Physics. Clearly, the way to
proceed is to study theoretically and experimentally polarized reactions
dominated by subprocesses with gluons in the initial state. To this effect,
experiments on charm production by polarized photons on longitudinally
polarized protons (polarized photoproduction) [1], large-$p_{T}$ direct
photon and jet production in polarized $p$-$p$ collisions [2], etc.
will be carried [3,4].

As a reaction leading to useful information, Ref. [5] has proposed the
production of large-$p_{T}$ hadron pairs $H_{1}$, $H_{2}$ in polarized photoproduction:
\[
\overrightarrow{\gamma }+\overrightarrow{p}\longrightarrow H_{1}+H_{2}+X
\]
An experiment could well be carried in COMPASS.

In view of this, we have undertaken an independent study
of the reaction (1.1). Working, as in [5], at Born level (leading order
(LO) in $\alpha_s$) we differ in the following from [5]:
\begin{itemize}
\item[(i)] We take into account the resolved photon contributions,
which are left out in [5].
\item[(ii)] In general, reaction (1.1) is dominated by the subprocesses 
\[
(a)\quad \overrightarrow{g} \overrightarrow{\gamma }\longrightarrow q
\overline{q},\qquad \;(b)\quad \overrightarrow{q} \overrightarrow{\gamma }%
\longrightarrow q g
\]
In [5], the first is well taken into
account, but the second is treated in a rather unclear way.
Here the subprocess (b) is treated on equal footing with (a).
\item[(iii)] We consider the effect of changing the renormalization
and factorization scales; in [5] this effect has also been left out.
\item[(iv)] In [5] the fragmentation of the final partons to hadrons
is treated via Monte-Carlo methods, which somewhat obscure the procedure.
Here we use the conventional QCD approach with recent fragmentation
functions [6].
\item[(v)] Very recently Ref. [7] presented an experimental result on (1.1);
this is discussed and compared to our predictions.
\item[(vi)] We show that a proper combination of cross sections
for certain choices of $H_{1}$ and $H_{2}$ will make a
more clean probe. The combination,
however, involves four cross sections, and the experiment will be more
difficult.
\end{itemize}

Furthermore, apart from the cross section calculated in [5] and in relation
with [7]
($\Delta d\sigma /d\phi_{1}dx$, for the definition of $\phi _{1}$
and $x$ see Sect. II), we present also results for the transverse momentum
distribution $\Delta d\sigma /d\phi _{1}dx_{T}$.

With the COMPASS experiment in mind (polarized muon-proton scattering),
we take into account that the (initial) photons are in general quasi-real
($\gamma^*$).

Sect. II presents our general formalism for the cross section $\Delta d\sigma
/d\phi _{1}dx$ and Sect. III for $\Delta
d\sigma /d\phi _{1}dx_{T}$. Sect. IV presents results for $\Delta d\sigma
/d\phi _{1}dx$ and the corresponding asymmetries. Sect. V
presents results for $\Delta d\sigma /d\phi _{1}dx_{T}$. Sect. VI presents
the above mentioned combination of cross sections as well as our results.
Finally, Sect. VII presents our concluding remarks.

\renewcommand{\theequation}{2.\arabic{equation}}
\setcounter{equation}{0}
\vglue 0.7cm
\begin{center}\begin{large}\begin{bf}
II. GENERAL FORMALISM FOR $\Delta d\sigma /d \phi _{1}dx$
\end{bf}\end{large}\end{center}
\vglue .2cm

The reaction (1.1) has, to some extent, been studied in Ref. [8], and
here we avoid repetition as much as possible. Consider the contribution of
the subprocess 
\[
\overrightarrow{a}\left( p_{1}\right) +\overrightarrow{b}\left( p_{2}\right)
\longrightarrow c_{1}\left( p_{3}\right) +c_{2}, 
\]
where the quantities in parentheses denote 4-momenta, and let 
\[
s=\left( p_{1}+p_{2}\right) ^{2},\qquad t=\left( p_{3}-p_{1}\right)
^{2},\qquad u=\left( p_{3}-p_{2}\right) ^{2} 
\]
($s+t+u=0$). Neglecting intrinsic transverse momenta, the hadrons $H_{i}$, ($
i=1,2$)$\ $are produced in opposite hemispheres with transverse momenta $
k_{iT}$ and c.m. pseudorapidities $\eta _{i}$ with respect to the photon.
Denoting by $\sqrt{S}$ the total c.m. energy and by $\phi _{1}$ the
azimuthal angle of $H_{1}$ and introducing 
\[
x_{iT}=2k_{iT}/\sqrt{S}, \non
\]
it follows that the cross section for (1.1) is formally given by [8,9] 
\[
\frac{\Delta d\sigma }{d\phi _{1}dx_{1T}dx_{2T}d\eta _{1}d\eta _{2}}=\frac{S
}{4}\int dx_{b}\Delta F_{b/\gamma }\left( x_{b}\right) \Delta \sigma \left(
S,x_{b},x_{1T},x_{2T},\eta _{1},\eta _{2}\right), 
\]
where $\Delta F_{b/\gamma }$ the polarized momentum distribution of parton $
b $ inside the photon and 
\[
\Delta \sigma =\frac{1}{\pi }\Delta F_{a/p}\left( x_{a}\right) \Delta \frac{
d\sigma }{d\widehat{t}}D_{H_{1}/c_{1}}\left( z_{1}\right)
D_{H_{2}/c_{2}}\left( z_{2}\right) ; 
\]
the limits of integration in (2.3) are specified later. In (2.4), $\Delta
d\sigma /d\widehat{t}$ is the cross section for the subprocess (2.1), $
D_{H_{i}/c_{i}}\left( z_{i}\right) $ is the fragmentation function for $
c_{i}\rightarrow H_{i}$ and 
\[
x_{a}=x_{b}\exp \left( -\eta _{1}-\eta _{2}\right) 
\]
\[
z_{i}=x_{iT}\left( \exp \left( \eta _{1}\right) +\exp \left( \eta
_{2}\right) \right) /2x_{b} 
\]
Eq. (2.4) expresses the contribution to the physical cross section from both
direct and resolved $\gamma $, the former corresponding to $\Delta
F_{b/\gamma }\left( x\right) =\delta \left( 1-x\right) $. The cross sections
for the subprocesses (1.2) are: 
\[
\Delta \frac{d\sigma_{g\gamma} }{d\widehat{t}}= -\frac{\pi aa_{s}e_{q}^{2}}{s^{2}%
}\frac{t^{2}+u^{2}}{tu},\qquad \Delta \frac{d\sigma_{q\gamma} }{d\widehat{t}} =%
\frac{8\pi aa_{s}e_{q}^{2}}{3s^{2}}\frac{s^{2}-t^{2}}{-st} 
\]
The corresponding cross sections for the resolved $\gamma $ contributions
are taken from [10] with $t\leftrightarrow u$ (see also [8]).

Here we define the variable 
\[
x=\exp \left( -\eta _{1}-\eta _{2}\right) 
\]
and determine first $\Delta d\sigma /d\phi _{1}dx$. Also introduce 
\[
h=\exp \left( \eta _{2}\right) 
\]
Taking, as in [8], the $x$-$z$ plane to be defined by $\overrightarrow{%
p_{2}}$ and $\overrightarrow{k_{1}}$ (i.e. $\phi _{1}=0$) we may write still
in a formal way 
\[
\frac{\Delta d\sigma }{d\phi _{1}dx}\left( S,\phi _{1}=0,x\right) = 
\frac{S}{4x}\int dx_{1T}\int dx_{2T}\int dx_{b}\Delta F_{b/\gamma }\left(
x_{b}\right)
\int \frac{dh}{h}\Delta \sigma \left(
S,x,x_{b},x_{1T},x_{2T},h\right) \non\\
\]
The physical meaning of the variable $x$ is clear from Eq. (2.5): it is $%
x=x_{a}$ for $x_{b}=1$ (direct $\gamma $).

The limits on $h$ are specified by the condition $z_{i}\leq 1$, which, in
view of (2.6), (2.8) and (2.9), implies: 
\[
h+x^{-1}h^{-1}\leq \lambda x_{b}, \non 
\]
where $\lambda \equiv \min \left( 2/x_{1T},2/x_{2T}\right) $. We find: 
\[
h_{-}\leq h\leq h_{+}, \non 
\]
where 
\[
h_{\pm }\equiv \frac{1}{2}\left[ \lambda x_{b}\pm \left( \lambda
^{2}x_{b}^{2}-\frac{4}{x}\right) ^{1/2}\right] ,\qquad h_{-}h_{+}=1/x 
\]
Clearly, we must have 
\[
\lambda x_{b}\geq 2/\sqrt{x} 
\]

Denoting the lower limit of $x_{1T}$, $x_{2T}$ integrations by $%
x_{T}^{\left( 0\right) }$ ($=2k_{T}^{\left( 0\right) }/\sqrt{S}$, $%
k_{T}^{\left( 0\right) }$ to be fixed by experiment) we write: 
\[
\int_{x_{T}^{\left( 0\right) }}^{x_{2T,max}}dx_{2T}=\int_{x_{T}^{\left(
0\right) }}^{x_{1T}}dx_{2T}+\int_{x_{1T}}^{x_{2T,max}}dx_{2T} 
\]
So we find: 
\[
\Delta \frac{d\sigma }{d\phi _{1}dx}\left( S,0,x\right) =\frac{S}{4}%
\int_{x_{T}^{\left( 0\right) }/\sqrt{x}}^{1}dx_{b}\left( I_{1}+I_{2}\right) 
\]
where 
\[
I_{1}=\int_{x_{T}^{\left( 0\right) }}^{x_{b}\sqrt{x}}dx_{1T}\int_{x_{T}^{%
\left( 0\right) }}^{x_{1T}}dx_{2T}\int_{h_{-}}^{h_{+}}\frac{dh}{h}\Delta
F_{b/\gamma }\left( x_{b}\right) \Delta \sigma \left(
S,x,x_{b},x_{1T},x_{2T},h\right) 
\]
with 
\[
h_{\pm }=x_{1T}^{-1}x_{b}\pm \left( x_{1T}^{-2}x_{b}^{2}-1/x\right) ^{1/2}, 
\]
and 
\[
I_{2}=\int_{x_{T}^{\left( 0\right) }}^{x_{b}\sqrt{x}}dx_{2T}\int_{x_{T}^{%
\left( 0\right) }}^{x_{2T}}dx_{1T}\int_{h_{-}}^{h_{+}}\frac{dh}{h}\Delta
F_{b/\gamma }\left( x_{b}\right) \Delta \sigma \left(
S,x,x_{b},x_{1T},x_{2T},h\right) 
\]
with 
\[
h_{\pm }=x_{2T}^{-1}x_{b}\pm \left( x_{2T}^{-2}x_{b}^{2}-1/x\right) ^{1/2} 
\]

Given $x_{T}^{\left( 0\right) }$, the condition (2.12) determines the
minimum value of $x$ allowable. Clearly
\[
x\geq \left( x_{T}^{\left( 0\right) }/x_{b}\right) ^{2} 
\]
and since $x_{b}\leq 1$:
\[
x_{\min} = \left( x_{T}^{\left( 0\right) }\right) ^{2} 
\]

\renewcommand{\theequation}{3.\arabic{equation}}
\setcounter{equation}{0}
\vglue 0.7cm
\begin{center}\begin{large}\begin{bf}
III. GENERAL FORMALISM FOR $\Delta d\sigma /d\phi _{1}dx_{T}$
\end{bf}\end{large}\end{center}
\vglue .2cm

We start again from Eqs. (2.3), (2.4) and change variables: 
\[
x_{2T}\rightarrow x_{T}=\frac{1}{2}\left( x_{1T}+x_{2T}\right) ,\qquad \eta
_{i}\rightarrow h_{i}=e^{\eta _{i}},\qquad i=1,2 
\]
so 
\[
\frac{\Delta d\sigma }{d\phi _{1}dx_{T}dx_{1T}dh_{1}dh_{2}}=\frac{S}{%
2h_{1}h_{2}}\int dx_{b}\Delta F_{b/\gamma }\left( x_{b}\right) \Delta \sigma
\left( S,x_{T},x_{b},x_{1T},h_{1},h_{2}\right) 
\]
and $\Delta \sigma $ given by (2.4). Now 
\[
x_{a}=x_{b}/h_{1}h_{2} 
\]
and 
\[
z_{1}=\frac{x_{1T}}{2x_{b}}\left( h_{1}+h_{2}\right) ,\qquad z_{2}=\frac{%
2x_{T}-x_{1T}}{2x_{b}}\left( h_{1}+h_{2}\right) 
\]
The conditions $z_{i}\leq 1$ and $x_{a}\leq 1$ imply 
\[
h_{2}+x_{b}h_{2}^{-1}\leq \lambda x_{b} 
\]
where $\lambda \equiv \min \left( 2/x_{1T},2/\left( 2x_{T}-x_{1T}\right)
\right) $. As in Sect. II: 
\[
h_{-}\leq h_{2}\leq h_{+} 
\]
where now 
\[
h_{\pm }\equiv \frac{1}{2}\left[ \lambda x_{b}\pm \left( \lambda
^{2}x_{b}^{2}-4x_{b}\right) ^{1/2}\right] ,\qquad h_{-}h_{+}=x_{b} 
\]
Here we have the condition 
\[
x_{b}\geq 4/\lambda ^{2} 
\]

Clearly, for $x_{1T}<x_{T}$: $\lambda =2/\left( 2x_{T}-x_{1T}\right) $,
whereas for $x_{1T}>x_{T}$: $\lambda =2/x_{1T}$. So in the present case we
write 
\[
\int_{x_{1T,min}}^{x_{1T,max }}dx_{1T}=\int_{x_{1T,min}}
^{x_{T}}dx_{1T}+\int_{x_T}^{x_{1T,max }}dx_{1T} 
\]
The final result is 
\[
\Delta \frac{d\sigma }{d\phi _{1}dx_{T}}\left( S,0,x_{T}\right) =\frac{S}{2}%
\left( J_{1}+J_{2}\right) 
\]
where 
\[
J_{1}=\int_{x_{1T,min
}}^{x_{T}}dx_{1T}\int_{x_{2T}^{2}}^{1}dx_{b}\int_{h_{-}}^{h_{+}}\frac{dh_{2}%
}{h_{2}}\int_{x_{b}/h_{2}}^{h_{1,max}}\frac{dh_{1}}{h_{1}}\Delta F_{b/\gamma
}\left( x_{b}\right) \Delta \sigma 
\]
with $x_{2T}\equiv 2x_{T}-x_{1T}$ and 
\[
x_{1T,min }=\max \left( x_{T}^{\left( 0\right) },2x_{T}-1\right) ,\qquad
h_{1,max }=2x_{b}/x_{2T}-h_{2} 
\]
\[
h_{\pm }=\frac{x_{b}}{x_{2T}}\pm \left( \frac{x_{b}^{2}}{x_{2T}^{2}}%
-x_{b}\right) ^{1/2}, 
\]
and 
\[
J_{2}=\int_{x_{T}}^{x_{1T,max
}}dx_{1T}\int_{x_{2T}^{2}}^{1}dx_{b}\int_{h_{-}}^{h_{+}}\frac{dh_{2}}{h_{2}}%
\int_{x_{b}/h_{2}}^{h_{1,max}}\frac{dh_{1}}{h_{1}}\Delta F_{b/\gamma }\left(
x_{b}\right) \Delta \sigma 
\]
with 
\[
x_{1T,max }=\min \left( 2x_{T}-x_{T}^{\left( 0\right) },1\right) ,\qquad
h_{1,max }=2x_{b}/x_{1T}-h_{2} 
\]
\[
h_{\pm }=\frac{x_{b}}{x_{1T}}\pm \left( \frac{x_{b}^{2}}{x_{1T}^{2}}%
-x_{b}\right) ^{1/2} 
\]
In determining $x_{1T,min }$ we took into account that $%
x_{1T}=2x_{T}-x_{2T}\geq 2x_{T}-1$, and in determining $x_{1T,max }$ that $%
x_{1T}=2x_{T}-x_{2T}\leq 2x_{T}-x_{T}^{\left( 0\right) }$.

\renewcommand{\theequation}{4.\arabic{equation}}
\setcounter{equation}{0}
\vglue .7cm
\begin{center}\begin{large}\begin{bf}
IV. RESULTS FOR $\Delta d\sigma /d\phi _{1}dx$ AND THE CORRESPONDING ASYMMETRIES
\end{bf}\end{large}\end{center}
\vglue .2cm

We present results for the three sets A,B,C of LO polarized distributions of 
[11], which can be roughly characterized as follows in terms of $\Delta
g\left( x\right) $ ($\equiv \Delta F_{g/p}\left( x,Q_{0}\right) $):
\begin{itemize}
\item[ ] Set A: $\Delta g\left( x\right) >0$ and relatively large
\item[ ] Set B: $\Delta g\left( x\right) >0$ and small
\item[ ] Set C: $\Delta g\left( x\right) $ changing sign; $\Delta g\left( x\right) <0$
for $x>0.1$.
\end{itemize}
The fragmentation functions $D_{H_{i}/c_{i}}$ are taken from [6] (LO
sets). In $a_{s}\left( Q\right) $ we use $\Lambda
=0.2$ $GeV$ and 4 flavors. The renormalization and factorization scales are
taken equal and with a central value $Q=Q_{c}\equiv k_{1T}+k_{2T}$.
We first present results at a typical COMPASS energy $\sqrt{S_{\gamma p}}\equiv
\sqrt{S}=12$ $GeV$ and for $k_{T}^{\left( 0\right) }=1.4$ $GeV$ [5].

Regarding the resolved $\gamma$ contributions, we have used the maximal
and minimal saturation sets of the polarized photon distribution functions
of [12,13]. We have also carried calculations with the distribution functions 
of [14], belonging to the class of the so-called asymptotic solutions,
and we simply report the results.

To account for the fact that the photons are quasi-real we multiply (2.4)
by the Weiszaecker-Williams factor:
\[
\Delta f\left( y \right) = \frac{\alpha}{2 \pi}
\Delta P_{\gamma l} \left( y \right) ln \frac{Q^2_{max} \left( 1-y \right) }
{m_{\mu}^2 y^2}
\]
where $\Delta P_{\gamma l} \left( y \right)=\left[ 1-\left( 1-y \right) ^2
 \right] / y$, $m_{\mu}=$ muon mass, and we take a typical value
$Q^2_{max}=4$ $GeV^2$; for incident lepton c.m. energy $\sqrt{S_l}$
(corresponding to $E_l=200$ $GeV$ [5]): $y=S/S_l$.

Fig. 1a presents $\Delta d\sigma /d\phi _{1}dx$ for direct and resolved $%
\gamma^* $ contributions with $H_{i}=\pi ^{+}$ or $\pi ^{-}$ (fragmentation
functions (A.4)-(A.8) of [6]).
The presented resolved contributions correspond to the maximal saturation
set of [12,13]; those of the minimal are somewhat smaller. So, in general,
the resolved contributions are much smaller than the direct.
However, in particular for the set A of [11] and in the range $%
0.15\leq x\leq 0.2$, where the direct contributions change sign,
the resolved are not insignificant. The asymptotic solution of [14]
gives even larger resolved contributions.

Notice that the differential cross sections for the direct $\gamma$
contributions change sign at some $x\leq 0.2$; this is due to the two
competing subprocesses of Eq. (1.2). At the lower $x$, $(a)$ dominates,
whereas at higher $x$, $(b)$ takes over. Hence the place to obtain
information about $\Delta G$ is at the lower $x$, as first was pointed
out in [5].  

Fig. 1b presents the asymmetries 
\[
A=\frac{\Delta d\sigma /d\phi _{1}dx}{d\sigma /d\phi _{1}dx} 
\]
for the sum direct+resolved and again $H_{i}=\pi ^{+}$ or $\pi ^{-}$. For the
unpolarized $d\sigma /d\phi _{1}dx$ we use the CTEQ distributions [15]
and the photon distribution functions of [16], LO sets.
Here, to account for quasi-real photons, in (4.1) we replace
$\Delta P_{\gamma l}$ by $P_{\gamma l} \left( y \right)=\left[ 1+\left( 1-y \right) ^2
 \right] / y$; hence the asymetry is reduced. 
For each of the sets A, B and C the strong line corresponds to the central
value $Q_{c}= k_{1T}+k_{2T}$. For the sets A and C, Fig. 1b presents also
the effect of changing the scales in the range $%
Q_{c}/2\leq Q \leq 2Q_{c}$.

Fig. 1b also presents an estimate of experimental
errors using the expression
\[
\delta A_{\gamma^* p}=\frac{1}{P_B P_T \sqrt{L \sigma_{\gamma^* p}} \epsilon}
\]
We take beam polarization $P_B=80\%$, target polarization $P_T=25\%$,
pion-kaon detection efficiency $\epsilon =1$
and integrated luminosity $L=2$ $fb^{-1}$ [5]; in (4.3)
$\sigma_{\gamma^* p}$ is the unpolarized cross section
for quasi-real photon-proton scattering integrated over a bin $\Delta x = 0.17$.

On the basis of Fig. 1b we conclude the following on the experiment:
First, the sets A and B cannot be distinguished. Second,
the sets A and C can barely be distinguished in the small range
$0.15\leq x \leq 0.2$. We note that at smaller $x$ the cross sections
$\gamma^* p$ become much smaller and $\delta A_{\gamma^* p}$ much
larger.

Now we turn to kaon production, and Fig. 2a presents $\Delta d\sigma /d\phi
_{1}dx$ for $H_{i}=K ^{+}$ or $K ^{-}$ (fragmentation functions (A.19)-(A.23) of
[6]) and $Q=Q_{c}$. The presented resolved contributions are as in Fig. 1a;
now for both sets A and B of [11], at $x \geq 0.25 $, they are important.
Fig. 2b presents the corresponding asymmetries together with the effect of
changing the scales and an estimate of the experimental errors, as for Fig. 1b.
Now the latter are significantly larger (smaller cross sections), making very
difficult the distinction even between sets A and C.

In [5], apart from kaons, the production of charged hadron pairs is considered.
The unpolarized cross sections for the production of charged hadrons are,
of course, greater than those of charged pions only. Thus the estimated errors will
be somewhat smaller.

As it has been stated, very recently an experimental result was presented
for (1.1). Its energy is low,
$\sqrt{S_l}=7.18$ $GeV$, and therefore $k_{iT}$ limited; also, the way
one reaches the final result is somewhat
unclear. Nevertheless, in view of its importance and of the fact that it is the
\textit{first experimental result}, it is perhaps of interest to extend the
calculation of Sects II and IV down to $\sqrt{S_l}=7.18$ $GeV$. 
This experiment selects events containing at least one positively
charged hadron and at least one negatively charged hadron. Hence the
fragmentation functions sould be separated to those for $\pi ^+$ $(K^+)$
and for $\pi ^-$ $(K^-)$. For the separation we use the subsequently
presented expressions (6.3)-(6.6). The fact that the experiment was carried
at a low energy necessitates the choice $k_T^{(0)}=1.1$ $GeV$. Taking into
account also the virtual photon depolarization factor $D=0.93$ [7], the 
\textit{predicted} asymmetries together with the experimental result are shown in
Fig 3; clearly, Set A (or B) is favored.

It is interesting also to note that the effect on the asymmetry of changing the
scales is small.   

The values $\sqrt{S}=12$ $GeV$ and $k_{T}^{\left( 0\right) }=1.4$ $GeV$ imply
certain limits on the rapidities
$\eta_i$ and the invariant mass of the hadron pairs $m(H_1H_2)$,
which amount to acceptance cuts. Taking the
variable $x$ in the range $0.055\leq x\leq 0.8$, the integration limits
for the variable $h$ in Eqs (2.15) and (2.17) combined with Eqs. (2.9) and
(2.8) imply the following limits on the rapidities:
$-1.9\leq \eta_1\leq 2.0$ and $-1.7\leq \eta_2\leq 2.1$.
These limits imply $m(H_1H_2)\geq 2.81$ $GeV$, which is in accord with [5].

It should be remarked that, instead of $Q_{c}= k_{1T}+k_{2T}$, the choice
$Q_{c}=\left( k_{1T}+k_{2T}\right) /2$ is also reasonable. Then varying
the scales in the range $Q_{c}/2\leq Q \leq 2Q_{c}$, near the lower
limit, with $k_{T}^0 =1.4$ $GeV$, we enter a region where perturbative
QCD is uncontrollable. One cannot take $k_{T}^0$ much larger because
$\Delta d\sigma /d\phi_{1}dx$ becomes too small.

\renewcommand{\theequation}{5.\arabic{equation}}
\setcounter{equation}{0}
\vglue .7cm
\begin{center}\begin{large}\begin{bf}
V. RESULTS FOR $\Delta d\sigma /d\phi _{1}dx_{T}$
\end{bf}\end{large}\end{center}
\vglue .2cm

The transverse momentum distributions $\Delta d\sigma /d\phi _{1}dx_{T}$ and 
$d\sigma /d\phi _{1}dx_{T}$ are calculated for the same distributions and
fragmentation functions as Sect. IV, as well as for $\sqrt{S}=12$ $GeV$ and $%
k_{T}^{\left( 0\right) }=1.4$ $GeV$. We present results only for $%
Q=Q_{c}\equiv k_{1T}+k_{2T}$ as functions of $x_{T}=\left(
k_{1T}+k_{2T}\right) /\sqrt{S}$.

The indicated errors have been estimated on the basis of Eq. (4.3) with the
unpolarized cross section integrated over a bin in $x_T$ corresponding to
$\Delta p_T =1$ $GeV$. 

Fig. 4a presents asymmetries for $H_{i}=\pi ^{+}$ or $\pi ^{-}$. Clearly, even
without accounting for the variation of the scales, sets A and C,
as well, are hard to distinguish.

Fig. 4b presents asymmetries for $H_{i}=K ^{+}$ or $K ^{-}$. 
The conclusions are the same as for Fig. 4a.

Again, as in Sect. IV, taking the variable $x_T$ in the range
$0.25\leq x_{T}\leq 0.8$ we obtain the following limits:
$-1.9\leq \eta_1\leq 2.1$, $-2.1\leq \eta_2\leq 2.1$ and
$m(H_1H_2)\geq 2.79$ $GeV$.

\newpage
\renewcommand{\theequation}{6.\arabic{equation}}
\setcounter{equation}{0}
\begin{center}\begin{large}\begin{bf}
VI. THE COMBINATIONS OF CROSS SECTIONS
\end{bf}\end{large}\end{center}
\vglue .2cm

Denote, for simplicity, $\sigma \left( H_{1}H_{2}\right) $ either of the
cross sections $\Delta d\sigma /d\phi _{1}dx$ and $\Delta d\sigma /d\phi
_{1}dx_{T}$ for $\overrightarrow{\gamma }+\overrightarrow{p}\longrightarrow
H_{1}+H_{2}+X$. As it is discussed in Refs. [17] and [8], neglecting the
resolved $\gamma $ contributions, the combinations
\[
\Delta \left( \pi \right) =\sigma \left( \pi ^{+}\pi ^{-}\right) +\sigma
\left( \pi ^{-}\pi ^{+}\right) -\sigma \left( \pi ^{+}\pi ^{+}\right)
-\sigma \left( \pi ^{-}\pi ^{-}\right) 
\]
and
\[
\Delta \left( K\right) =\sigma \left( K^{+}K^{-}\right) +\sigma \left(
K^{-}K^{+}\right) -\sigma \left( K^{+}K^{+}\right) -\sigma \left(
K^{-}K^{-}\right) 
\]
isolate the contribution of the subprocess $\overrightarrow{g}%
\overrightarrow{\gamma}\longrightarrow q\overline{q}$.

When the resolved $\gamma $ contributions, calculated via the polarized
distribution functions of [12,13] or [14] are taken into account, the contribution of $%
\overrightarrow{q}\overrightarrow{\gamma }\longrightarrow qg$ is not
completely eliminated, but we find that for $x\leq 0.4$ it is smaller by $%
\sim 2$ orders of magnitude than the contribution of $\overrightarrow{\gamma 
}\overrightarrow{g}\longrightarrow q\overline{q}$. Hence the difference in $%
\Delta g$ between the sets A, B and C is displayed much better. Below we
present results for the corresponding asymmetries and for $Q=Q_{c}\equiv
k_{1T}+k_{2T}$.

The calculation of (6.1) requires the separation of the fragmentation
functions for $\pi ^{+}$ and $\pi ^{-}$. To this purpose, as in [8], we use:
\[
D_{\pi ^{+}/u}\left( z\right) /D_{\pi ^{-}/u}\left( z\right) =D_{\pi
^{-}/d}\left( z\right) /D_{\pi ^{+}/d}\left( z\right) =\frac{1+z}{1-z}
\]
and
\[
D_{\pi ^{+}/g}\left( z\right) /D_{\pi ^{-}/g}\left( z\right) =D_{\pi
^{+}/s}\left( z\right) /D_{\pi ^{-}/s}\left( z\right) =1
\]
For the calculation of (6.2) we use [8]:
\[
D_{K^{+}/u}\left( z\right) /D_{K^{-}/u}\left( z\right) =D_{K^{+}/\overline{s}%
}\left( z\right) /D_{K^{-}/\overline{s}}\left( z\right) =\frac{1+z}{1-z}
\]
and
\[
D_{K^{+}/g}\left( z\right) /D_{K^{-}/g}\left( z\right) =D_{K^{+}/d}\left(
z\right) /D_{K^{-}/d}\left( z\right) =1
\]

The effect of changing the scales is very similar to that of Figs
1b and 2b.

The indicated errors have been estimated as follows: First,
for each of the cross sections $\sigma \left( H_{1}H_{2}\right) $
we have determined an error (via Eq. (4.3)) and then we have taken
the square root of the sum of the squares of these errors (assuming
independent measurments of $\sigma \left( H_{1}H_{2}\right) $).

Fig. 5 displays asymmetries corresponding to cross sections $\Delta d\sigma
/d\phi _{1}dx$; Fig. 5a refers to $\Delta \left( \pi \right) $ and 5b to $%
\Delta \left( K\right) $. Now the differences between the sets A, B and in
particular C are larger and over a wider range of $x$ than
in Figs. 1b and 2b. The errors are larger, but $\Delta \left( \pi \right) $
appears to be useful in distinguishing between sets A and C. As for
$\Delta \left( K \right) $, the errors are too large to be of any use.

Fig. 6 displays asymmetries corresponding to cross sections $\Delta d\sigma
/d\phi _{1}dx_{T}$; Fig. 6a refers to $\Delta \left( \pi \right) $ and 6b to 
$\Delta \left( K\right) $. Again, the differences between sets A, B and in
particular C are larger. Fig. 6a seems to show that $\Delta \left( \pi \right) $
does distinguish between sets A and C at $0.25 \leq x \leq 0.3$.
Again, for $\Delta \left( K \right) $ the errors are too large.

Of course, as in Sects IV and V, we present predictions for $\pi$
and $K$ separetely. In an experiment detecting $\pi +K$ the errors will be
somewhat smaller.

\renewcommand{\theequation}{7.\arabic{equation}}
\setcounter{equation}{0}
\vglue .7cm
\begin{center}\begin{large}\begin{bf}
VII. CONCLUDING REMARKS
\end{bf}\end{large}\end{center}
\vglue .2cm

On the basis of Fig 1b we have concluded that the sets
A and C of polarized parton disrtibutions can barely be distinguished
and only in the small range $0.15\leq x\leq 0.2$.
The sets A and B cannot be distinguished.

Nevertheless, as we stated, in an experiment detecting all charged particles,
the errors will be smaller. Furthermore, one way to increase the asymmetry (4.2)
is by increasing $\sqrt{S}=\sqrt{S_{\gamma p}}$; this will somewhat reduce
the implications of producing quasi-real instead of real photons. Alternatively, one may decrease
the incident lepton c.m. energy $\sqrt{S_{lp}}$. On the other hand, producing
quasi-real photons with $\sqrt{S}$ very near $\sqrt{S_{lp}}$ might
make the experiment more difficult [8].

Polarized real photons at energy comparable to that of the
present paper $\sqrt{S}$ ($\simeq 10$ $GeV$) are available at SLAC.

A very recent experiment at $\sqrt{S}=7.18$ $GeV$ [7] favors set A or B (Fig. 3).
However, the fact that this energy is low and the way the final result is
obtained makes necessary the repetition of the experiment at a higher energy [1] as well as
experiments at even higher energies with different reactions involving
polarized initial particles [2].
 
A somewhat better probe of $\Delta g$ appears at first sight to be the combination
$\Delta \left( \pi \right) $ of cross sections corresponding to
$\Delta d\sigma /d\phi_{1}dx$ (Fig. 5a) and even better the combination
$\Delta \left( \pi \right) $ corresponding to
$\Delta d\sigma /d\phi_{1}dx_{T}$ (Fig. 6a). However, in our estimate of errors,
only statistical ones are taken into account. The systematic errors in an
experiment measuring four cross sections may be significant; this holds even
more if experiments at different places are involved. 
 
In this work (and in [5]) the effect of next-to-leading order corrections
(NLOC) has not been considered. A number of other cases suggests that their
effect on the asymmetries will be less important than that on the cross
sections; a partial understanding can be found in Ref. [18]. With NLOC, the
effect on the cross sections of changing the scales is, in general, reduced.
Whether (and how much) this affect will be reduced on the asymmetries is
unclear. 
Unclear also is to what extend NLOC affect the combinations
$\Delta \left( \pi \right) $ and $\Delta \left( K \right) $,
which at LO isolate the subprocess
$\overrightarrow{g} \overrightarrow{\gamma }\longrightarrow q
\overline{q}$.
Anyway, the interest in reaction (1.1) as a possible probe of $%
\Delta g$ makes imperative the determination of NLOC.

\newpage 
\begin{center}\begin{large}\begin{bf}
ACKNOWLEDGMENTS
\end{bf}\end{large}\end{center}
\vglue .2cm

We would like to thank D. de Florian and W. Vogelsang for useful discussions,
D. von Harrah for giving us a preprint of Ref. 5 and pointing to us this problem,
and M. Svec for checking certain of our results.  Thanks are due also for discussions
to a number of colleagues, in particular Gerry Bunce, at Brookhaven National Lab,
where one of us (APC) spent some time during September 1999.
The work was also supported by
the Secretariat for Research and Technology of Greece and by the Natural Sciences
and Engineering Research Council of Canada. 

\vglue 2.5cm
\begin{center}\begin{large}\begin{bf}
REFERENCES
\end{bf}\end{large}\end{center}
\vglue .3cm

   \begin{list}{$[$\arabic{enumi}$]$} 
    {\usecounter{enumi} \setlength{\parsep}{0pt} 
     \setlength{\itemsep}{3pt} \settowidth{\labelwidth}{(99)} 
     \sloppy} 
\item
G.Baum et al., COMPASS Collaboration,
CERN/SPSLC-96-14, CERN/SPSLC-96-30.
\item
D. Hill et al., RHIC Spin Collaboration,
"Proposal on Spin Physics" (update), 1993; Proccedings of RIKEN BNL Research Center Workshop,
BNL-65615, April 1998.
\item
B. Lampe and E. Reya,
preprint MPI-PhT/98-23 and DO-TH 98/02.
\item
M. Anselmino, E. Efremov and E Leader,
Phys. Rep. 261, 1 (1995); H.-Y. Cheng,
Int. J. Mod. Phys. A11, 5109 (1996).
\item
A. Bravar, D. von Harrah and A. Kotzinian,
Phys. Lett. B421, 349 (1998).
\item
J. Binnewies, B Kniehl and G. Kramer,
Phys. Rev. D52, 4947 (1995).
\item
A. Airapetian et al (HERMES Collab.), hep-ex/9907020
\item
J. Peralta, A. P. Contogouris, B. Kamal and F. Lebessis,
Phys. Rev. D49, 3148 (1994).
\item
S. D. Ellis and M. Kislinger,
Phys. Rev. D9, 2027 (1974).
\item
J. Babcock, E. Monsay and D. Sivers,
Phys. Rev. D19, 1483 (1979).
\item
T. Gehrman and W. Stirling,
Phys. Rev. D53, 6100 (1996).
\item
M. Glueck and W. Vogelsang,
Z. Phys. C55, 353 (1992); C57, 309 (1993).
\item
M. Glueck, M. Stratmann and W. Vogelsang,
Phys. Lett. B337, 373 (1994).
\item
J. Hassan and D. Pilling,
Nucl. Phys. B187, 563 (1981).
\item
H. Lai et al.,
hep-ph/9903282
\item
M. Glueck, E. Reya and G. Vogt,
Phys. Rev. D46, 1973 (1992).
\item
M. Fontannaz, B. Pire and D Schiff,
Z. Phys. C6, 563 (1981).
\item
A. P. Contogouris, S. Papadopoulos and F. Tkachov,
Phys. Rev. D46, 2846 (1992);
A. P. Contogouris, B. Kamal, Z. Merebashvili and F. Tkachov,
ibid D48, 4092 (1993); D54, 7081 (1996) (E).
\end{list}

\newpage
\begin{center}\begin{large}\begin{bf}
FIGURE CAPTIONS
\end{bf}\end{large}\end{center}

\begin{list}{Fig.~\arabic{enumi}.} 
    {\usecounter{enumi} \setlength{\parsep}{0pt} 
     \setlength{\itemsep}{3pt} \settowidth{\labelwidth}{Fig.~9.} 
     \sloppy } 
\item 
Results when each of the final hadrons $H_{i}$, $i=1,2$,
is $\pi ^{+}$ or $\pi ^{-}$.

(a) Differential cross sections $\Delta d\sigma /d \phi _{1}dx$ for
direct and resolved $\gamma $ contributions for $Q=Q_{c}= k_{1T}+k_{2T}$.
A, B and C refer to the parton distributions of Ref. [9].
 
(b) Asymmetries $A=\left( \Delta d\sigma /d\phi _{1}dx\right) / \left( d\sigma /d\phi _{1}dx \right)$
and their variation with changing the scales in the range
$Q_{c}/2\leq Q \leq 2Q_{c}$.
Strong lines correspond to the scale $Q=Q_c$. The bands with forward and backward
slanted hatches show this variation for sets A and C correspondingly.
For set B the variation is not shown.
\item 
The same as in Fig. 1 when each of the final hadrons $H_{i}$
is $K^{+}$ or $K^{-}$.
\item
The predicted asymmetries at  $\sqrt{S}=7.18$ $GeV$ together
with the recent experimental result of Hermes collaboration [7].
\item 
Asymmetries $A=\left( \Delta d\sigma /d\phi _{1}dx_T \right) /
\left( d\sigma /d\phi _{1}dx_T \right) $.

(a) When each of the hadrons $H_{i}$ is $\pi ^{+}$ or $\pi ^{-}$.

(b) When each of $H_{i}$ is $K^{+}$ or $K^{-}$.
\item 
Asymmetries for the combinations (6.1) and (6.2) for cross sections
$\Delta d\sigma /d\phi _{1}dx $.

(a) For the combination (6.1) of pions

(b) For the combination (6.2) of kaons
\item 
The same as in Fig. 4, but referring to the cross sections $\Delta d\sigma /d\phi _{1}dx_T$  
\end{list}

\end{document}